\documentclass[preprint,english,showpacs,11pt,floatfix]{revtex4}
\usepackage{babel,amsmath,amssymb,dcolumn}
\usepackage[dvips]{graphics}
\usepackage{hyperref}

\usepackage{amsfonts}
\usepackage{amssymb}
\usepackage{latexsym}
\usepackage[pdftex]{graphicx}
\usepackage{dcolumn}

\newcommand{\be}{\begin{equation}}
\newcommand{\ee}{\end{equation}}

\newcommand{\bear}{\begin{eqnarray}}
\newcommand{\eear}{\end{eqnarray}}

\begin{document}

\title{Signature of the interaction between dark energy and
dark matter in observations}

\author{Elcio Abdalla}
\email{eabdalla@fma.if.usp.br}
\author{L. Raul Abramo}
\email{abramo@fma.if.usp.br}
\author{Jos\'e C. C. de Souza}
\email{jccampos@fma.if.usp.br}
\affiliation{Instituto de F\'isica, Universidade de S\~ao Paulo, C.P.
  66318, 05315-970, S\~ao Paulo, SP, Brazil}

\begin{abstract}
  We investigate the effect of an interaction between dark energy and
  dark matter upon the dynamics of galaxy clusters.  This effect is
  computed through the Layser-Irvine equation, which describes how an
  astrophysical system reaches virial equilibrium and was modified to
  include the dark interactions. Using observational data from almost
  100 purportedly relaxed galaxy clusters we put constraints on the
  strength of the couplings in the dark sector. We compare our results
  with those from other observations and find that a positive (in the
  sense of energy flow from dark energy to dark matter) non vanishing
  interaction is consistent with the data within several standard
  deviations.
\end{abstract}


\maketitle

\section{Introduction}

One of the most surprising results of physics and cosmology in the
last ten years is the cosmological accelerated expansion, which has
been proved beyond reasonable doubt by observations
\cite{cosmoaccel,wmap,bao}. The simplest explanation of such
observations is a cosmological constant, which is, however, off
theoretical computations by 120 orders of magnitude, a result that
calls for urgent explanations. Another possibility is to argue
that the Universe contains a strange dynamical component with negative
pressure, dark energy, which is responsible for more than three
quarters of its energy content \cite{darkenergy}.

Interactions of dark energy or dark matter with baryonic matter and
radiation must be either inexistent or negligible in order to comply
with stringent observations of the visible sector.  However, in a
field-theoretical approach, dark matter is some particle of a
unification scheme, and dark energy should also be described by a
field, which means that some level of interaction between these
different sectors is basically mandatory. In the literature there is a
large body of work dealing with such a possibility, e.g.,
\cite{amendola}-\cite{[14]}. It is interesting to notice that a
coupling between dark energy and dark matter can also serve to
alleviate the coincidence problem \cite{amendola,[11]}. The value of
the coupling and holographic arguments also allow for the crossing of
the phantom barrier which separates models with equations of state $w
\equiv p/\rho > -1$ from models with $w < -1$ \cite{[12]} -- see also
\cite{[14],[15]}. Further consequences of the interaction have been
studied, as {\it e.g.} its effect on the lowest multipoles of the
cosmic microwave background (CMB) spectrum \cite{[13],[16]}.  The
strength of the coupling is presumably as small as the fine structure
constant \cite{[13],[17]}. Comparison with supernova data and CMB and
large-scale structure have also been analysed \cite{[18]}.
Nevertheless, the observational limits on the strength of such an
interaction remain weak \cite{[19]}.

In addition to these constraints,
dynamical dark energy (i.e., a time- and space-dependent field)
has an impact on the large-scale structure due to its
fluctuations. In that case, dark energy affects the process of
structure formation by means of its density fluctuations, both in the
linear \cite{[11]}, \cite{[20]}-\cite{[23]} and the non-linear
\cite{[24],[25]} regimes, as the growth of dark matter perturbations
can be affected \cite{[13],[14],[26]}.

However, most constraints on dark interactions
concern the asymptotic  behaviour of the dark
sector, both in time and in space.
Local or nearby checks are still weak. This started to change
with the detailed analysis of galaxy clusters and their
internal structure.
It was suggested that the dynamical equilibrium of collapsed
structures could be affected by the coupling of dark energy to dark
matter, which as a result could affect the averaged energy distributions
predicted by the virial theorem -- in fact, by
its relativistic counterpart, the Layser-Irvine
equation \cite{[28]}. This was first proposed, and analyzed for the
Abell cluster A586, by \cite{[27]}.  The basic idea is that the virial
theorem is distorted by the mass non-conservation generated by
the coupling of dark matter with dark energy.

In a previous paper \cite{ababrasw} we showed how the Layser-Irvine
equation, describing the evolution towards virialization, is changed by the
presence of the coupling. We showed that this violation leads to a
systematic bias in the estimation of virial masses of clusters when
the usual virial conditions are employed.
By comparing weak lensing and X-ray mass-observables
on the one hand, and virial masses on the other hand, we were able to
cross-check not only the strength of the coupling, but also whether it
was indeed a feature of the virial mass estimate, and not a more
complicated set of biases between all these mass-observable relations.
Our main result was that a single parameter could explain all
the bias between virial mass and the other estimates (weak lensing and
X-ray) to 95\% confidence level (C.L.), or, equivalently, two standard
deviations.

However, the amount of data (about thirty independent useful
observations) did not allow a detailed numerical analysis, resulting
in relatively weak constraints on the coupling.
In this paper we will not only reanalyze that data, but will also
include a new set of data for almost a hundred X-ray
observations of galaxy clusters, and compare the presumed
mass from those observations with the virial masses for those clusters.

Notice that even though the uncertainties associated with any
individual galaxy
cluster are very large, by comparing the naive virial masses of a
large sample of clusters with their masses estimated by X-ray and by
weak lensing data, we are able to constrain the bias between them
and to impose much tighter limits on the strength of the coupling
than has been achieved before.

\section{Phenomenology of Coupled Dark
Energy And Dark Matter Models}

We start with a simplified two fluid model interaction:
\begin{eqnarray}
\dot\rho_{dm} + 3 H \rho_{dm} &=& \psi
\\ \nonumber
\dot\rho_{de} + 3 H \rho_{de} (1+w_{de}) &=& - \psi \; ,
\end{eqnarray}
where a dot denotes time derivative, $H$ is the expansion rate,
$\rho_{dm}$ and $\rho_{de}$ are respectively the energy densities of
dark matter and dark energy, and $w_{dm}$ and $w_{de}$ are their
equations of state. Notice that the continuity equation still holds
for the total energy density $\rho_{Tot} = \rho_{dm} + \rho_{de}$.

Phenomenologically, one can describe the interaction between the two
fluids as an exchange of energy at a rate proportional to a
combination of the energy densities \cite{[8],[12]}:
\be
\label{psi}
\psi = \zeta H \rho_{Tot} \; .
\ee
In fact, the term proportional to $\rho_{dm}$ could lead
to instabilities \cite{maartens,he}, but that does not concern us here
in view of the local character of the computation. We are interested
in collapsed structures where the local, inhomogeneous density
$\sigma$ is far from the average, homogeneous density $\rho$. In that
case the continuity equation for dark matter reads \cite{[25]}:
\be
\label{cont}
\dot\sigma_{dm} + 3H\sigma_{dm} +
\vec\nabla \left( \sigma_{dm} \vec{v}_{dm} \right) =
\zeta H \left( \sigma_{dm} + \sigma_{de} \right) \; ,
\ee
where $\vec{v}_{dm}$ is the peculiar velocity of dark matter particles.
We have considered the local density of dark
energy to be proportional to the local density of dark
matter, $\sigma_{de} = b_{em} \sigma_{dm}$.
If for a given model the dark energy component is very homogeneous,
then $b_{em} \approx 0$.
We do not consider the case where $b_{em}$ depends on
the size and mass of the collapsed structure -- although
this should probably
happen in realistic models of structure formation with dark
energy \cite{[25]}. Hence, the continuity equation with dark
matter coupled to dark energy reads:
\be
\label{cont_eq}
\dot\sigma_{dm} + 3H\sigma_{dm} +
\vec\nabla \left( \sigma_{dm} \vec{v}_{dm} \right) =
\bar\zeta H \sigma_{dm} \; ,
\ee
where $\bar\zeta = \zeta (1+b_{em})$.

Considering that the kinetic ($ K_{dm}$) and potential ($U_{dm}$)
energies of a set of particles interact via gravity and the coupling
(\ref{psi}), we found that the corrected Layser-Irvine equation leads
to the condition:
\be
\label{LI}
(2-\bar\zeta) K_{dm} + (1+b_{em}) (1-2\bar\zeta) U_{dm} = 0 \; .
\ee
Taking $\bar\zeta = b_{em} = 0$ we recover the usual virial condition.

\section{Mass estimation and limits on the coupling}

In order to find deviations from the virial theorem we analyse galaxy
clusters, which are the largest virialized structures in the Universe
\cite{Ettori,Bahcall}. Studies of cluster in the optical,
in X-ray and through weak lensing are available in the literature,
which can then be used to
estimate the mass of the cluster, under some hypotheses.

Clusters can be observed directly in the optical wavelengths, where the
radial velocities are estimated through their projected (line-of-sight)
velocities. The velocity field, together with the projected
spatial distribution of galaxies in the cluster, allow for an estimate
of the relative shares of the potential and kinetic energy of its constituents.
Assuming that the clusters are virialized, their
masses can then be computed. But if there is a coupling of dark energy
to dark matter, we find that \cite{ababrasw}
\be
\label{equil}
(1+b_{em}) \frac{U_{dm}}{K_{dm}}
= -2 \frac{1-\bar\zeta/2}{1-2\bar\zeta} \; .
\ee
Since the potential energy is proportional to the square of the
masses, and the kinetic energy is only linearly proportional to
the mass, there is a shift in the usual virial mass estimation, which
is entirely due to the coupling, by a factor of
$ (1-2\bar\zeta)/(1-\bar\zeta/2)$.

One can measure the mass
with other observations by making very simple assumptions which
have nothing to do with the precise nature of the equilibrium
of the system.
In particular, both the weak lensing and X-ray methods
provide physical observables which can be used to
estimate the total mass of a cluster in ways which are totally
independent of the virial method.

In what follows we will assume that the variations in the
mass estimators can be explained by a single parameter -- the
coupling between dark energy and dark matter. Of course, all
these mass estimates are rife with systematic errors of many types,
and intrinsic uncertainties in these estimates should be expected.
For instance, if the gas in a cluster is not homogeneously
distributed, the clumps of gas will enhance the X-ray luminosity
of the cluster relative to a homogeneous distribution
(as the X-ray luminosity is proportional to the
square of the number density of free electrons), leading to
a possible overestimation of the X-ray mass.
Clearly, our method cannot make a distinction between the
constraint on the coupling of the dark sector and some other
source of systematic error, such as the clumpiness of gas in
clusters.

We parametrize this single-parameter deviation in mass estimations,
both with weak lensing or X-ray data, as:
\be
\label{Mvir}
M_{vir}
= \frac{1-\bar\zeta/2}{1-2\bar\zeta} M_{X}
= \frac{1-\bar\zeta/2}{1-2\bar\zeta} M_{WL}
\ee
Previously \cite{ababrasw}, we compared and cross-checked
the three ratios that can be obtained from each pair of masses.
We obtained that $\zeta = 0.04\pm 0.02$, and checked that there was
no bias between X-ray and weak lensing estimates.

The data in \cite{girardif} allows to more than triple our dataset,
compared with what we had
before, leading to the possibility of a more thorough numerical analysis.
First, we can achieve a better accuracy, showing that
the coupling is non zero by several standard deviations.
And second, we also
analyzed a possible dependence of the coupling on the size of the
cluster as well as on its redshift -- checking for possible evolution effects.
Hence, since many more X-ray data for clusters are available than for weak
lensing, we can enhance our test
of the dark interaction by including X-ray clusters without
the weak lensing counterpart.

\begin{figure}[!htp]
\begin{center}

\includegraphics[width=7cm]{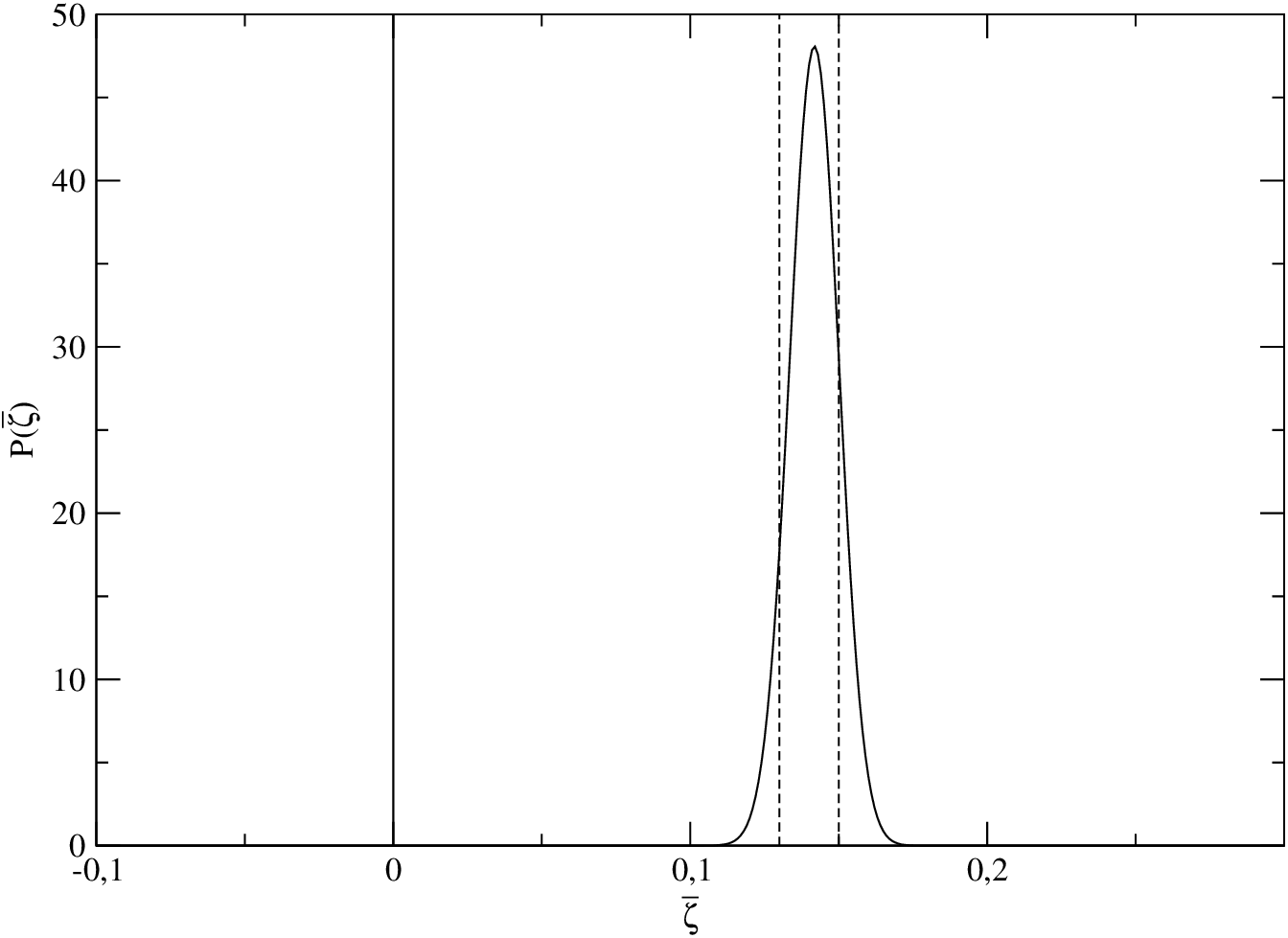}
\caption{Probability density function (p.d.f.) for $\bar\zeta$ obtained with
  all available cluster masses.
  The result points to $\bar\zeta = 0.14\pm 0.01$, that
  is, 14 standard deviations from the null result.}
\label{figurejoint}

\end{center}
\end{figure}

We can collect all these datasets by taking each mass estimate,
together with its uncertainties, and constructing a likelihood
function.  Our data comprises a set of 34 weak lensing, X-ray and
virial mass estimates from \cite{Cypriano1,Cypriano2,Hoekstra}, as
well as 61 X-ray and virial (optical) mass estimates from
\cite{girardif}. The weak lensing and X-ray masses were compared with
the optical data, and the results are represented in figure
(\ref{figurejoint}).

One of the difficulties, already encountered in our previous paper, is
how to deal with asymmetric errors. At that time we just worked with
the average error, discarding corrections due to such an asymmetry.
Taking this asymmetry into account, e.g. by making use of the
techniques presented in Ref. \cite{barlow}, may lead to more accurate
constraints, however, there are some difficulties associated with the
asymptotic behaviour of the likelihood function, and the final results
are not significantly different from the ones obtained with a more
simple analysis. Hence, we employ the simpler, symmetric treatment of
the errors.

With the 61 X-ray  masses mentioned above
(including, of course, the optical data), as well as the earlier 34
X-ray and weak lensing masses, we can construct
the likelihood function $\cal{L}$ for $\bar\zeta$ with the full dataset:
\begin{equation}
\label{Like}
{\cal{L}} \propto
\prod_{i=1}^{N} \exp
\left\{  - \frac{1}{2 \, \sigma^2_i}
\left[
\frac{1-2\bar\zeta}{1-\bar\zeta/2} - f_{i}
\right]^2
\right\} \; ,
\end{equation}
where $f$ represents the mass ratios $M_X/M_{vir}$ and
$M_{WL}/M_{vir}$, and the errors $\sigma$ are approximated by
a simple geometric mean of the asymmetric uncertainties for each of
the data, $\sqrt{\sigma_{+}\sigma_{-}}$. Setting a conservative prior for the
values of $\bar\zeta$ to be flat, and between -0.3 and 0.3, we obtain the
probability density function displayed in figure (\ref{figurejoint}). We
learn that X-ray data are more robust. The weak lensing data are more
spread, though consistent with X-ray measurements \cite{ababrasw}.

\vspace{0.5cm}

The joint analysis point to:
\be
\bar\zeta = 0.14 \pm 0.01\quad .\label{joint}
\ee

Notice that the likelihood function assumes implicitly that all the
variation in the mass estimations can be attributed to a single
parameter, and the peak and width of that likelihood give us the
best-fit and the uncertainty for that parameter. We could also have
made no assumption whatsoever about the nature of these estimates,
and compute the sample variance of all mass estimates. The sample
variance of the ratio between mass estimates (which is, again,
degenerate with our coupling parameter) results, of course,
in a much higher variance, namely $\sigma^2=0.3$.

The main result obtained here is that, assuming that a single
parameter can explain the difference between the mass estimations,
there is a positive coupling which is nonzero within several
standard deviations. Now, in order to investigate whether different
masses or evolution effects, for example,
could be playing a role in our analysis,
we can divide our data into subsets.

\begin{figure}[!htb]
\centering
\includegraphics[scale=0.21]{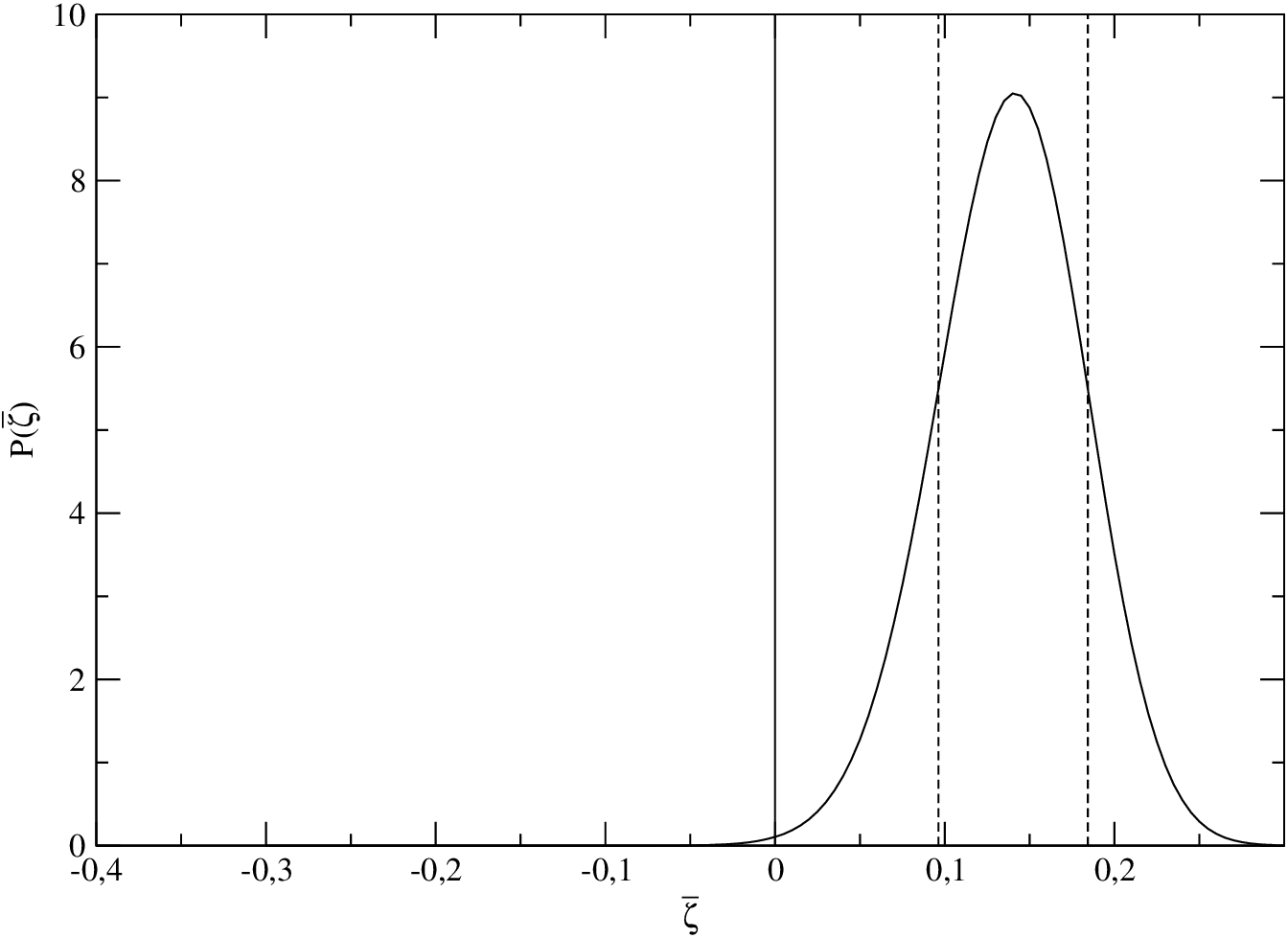}
\hspace{0.5cm}
\includegraphics[scale=0.21]{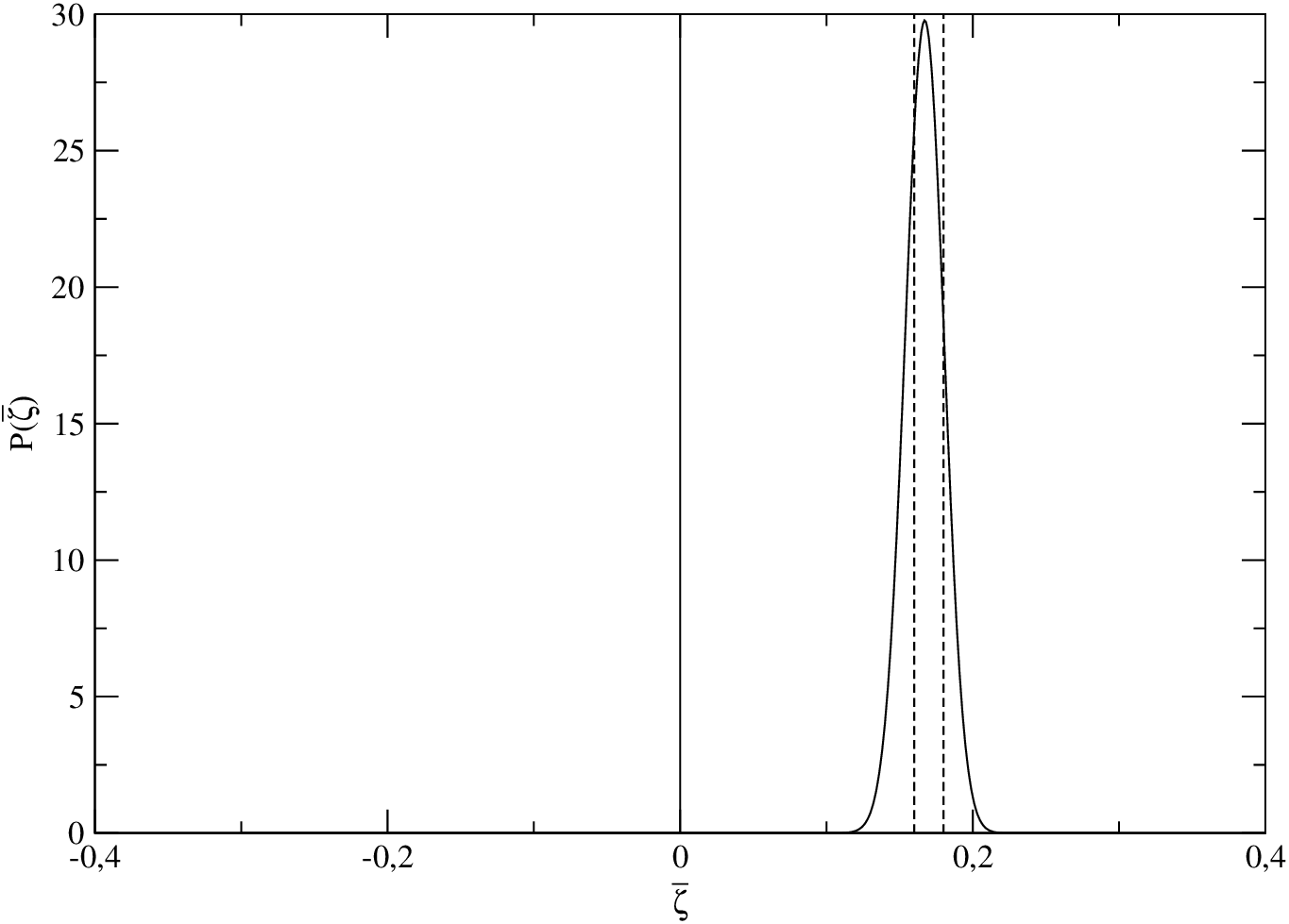}
\hspace{0.5cm}
\includegraphics[scale=0.21]{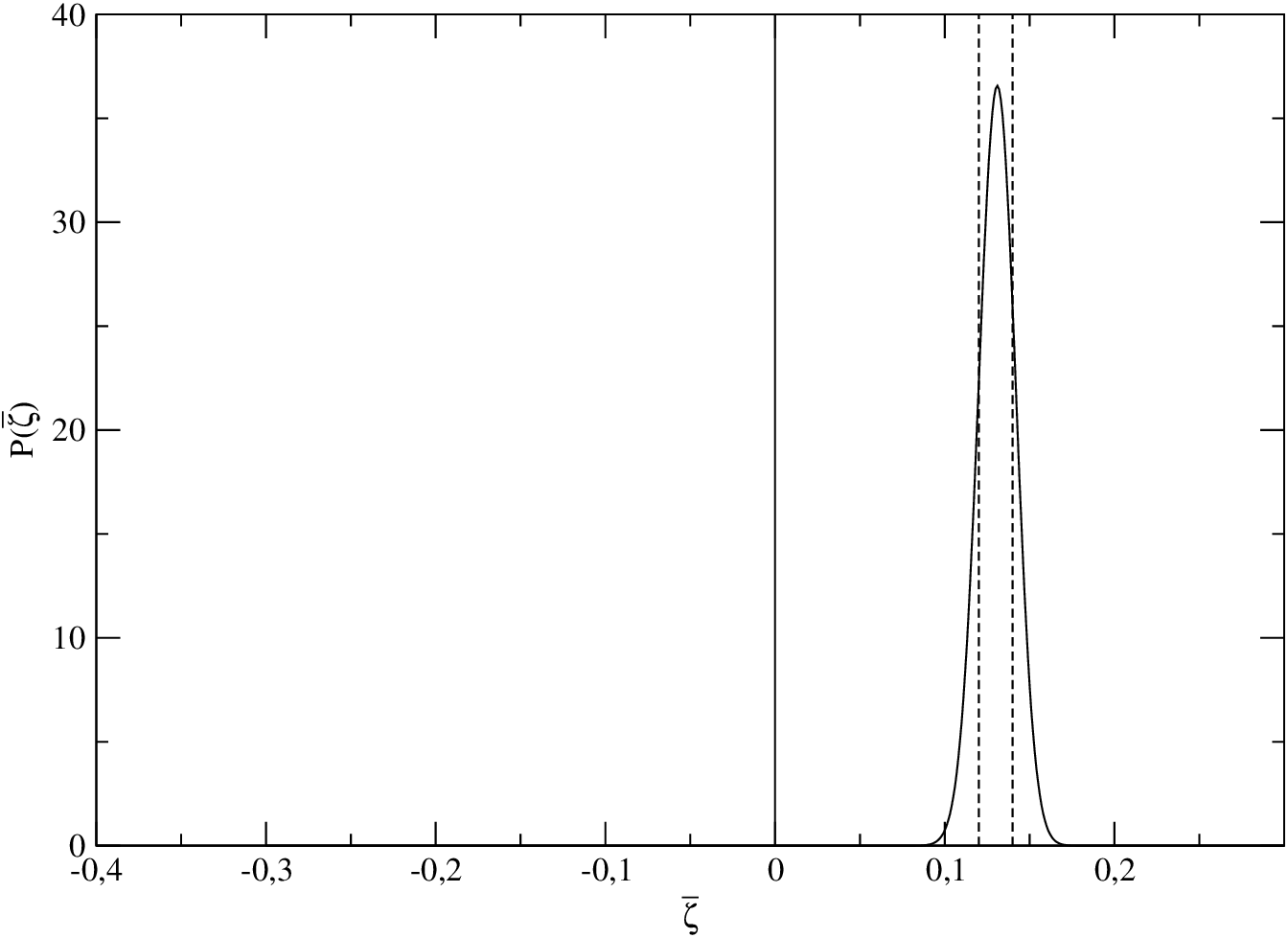}
\caption{P.d.f's concerning the cluster data divided in three sets, in
  increasing order of the dispersion velocity (up to 600 km/s, from
  600 to 1000 km/s, and above 1000 km/s.) The data are compatible with one
  another and the results point to essentially the same values of $\bar\zeta$,
  all at least three standard deviations from the null result.}
\label{figure3dia}
\end{figure}

First, we can redo the numerical analysis
after dividing the data in terms of the velocity
dispersion -- which is in fact a proxy for mass.
This analysis leads to the three diagrams shown in figure
(\ref{figure3dia}).

\vspace{0.5cm}

The results for $\bar\zeta$ are, respectively, for smaller to larger
values of the velocity dispersion:
\bear
\bar\zeta_{v_{1}} &=& 0.14 \pm 0.04 \nonumber\\
\bar\zeta_{v_{2}}&=& 0.17\pm 0.01 \label{3diaresults}\\
\bar\zeta_{v_{3}}&=&0.13\pm 0.01 \; .
\nonumber
\eear

\begin{figure}[!htp]
\begin{center}
\includegraphics[scale=0.21]{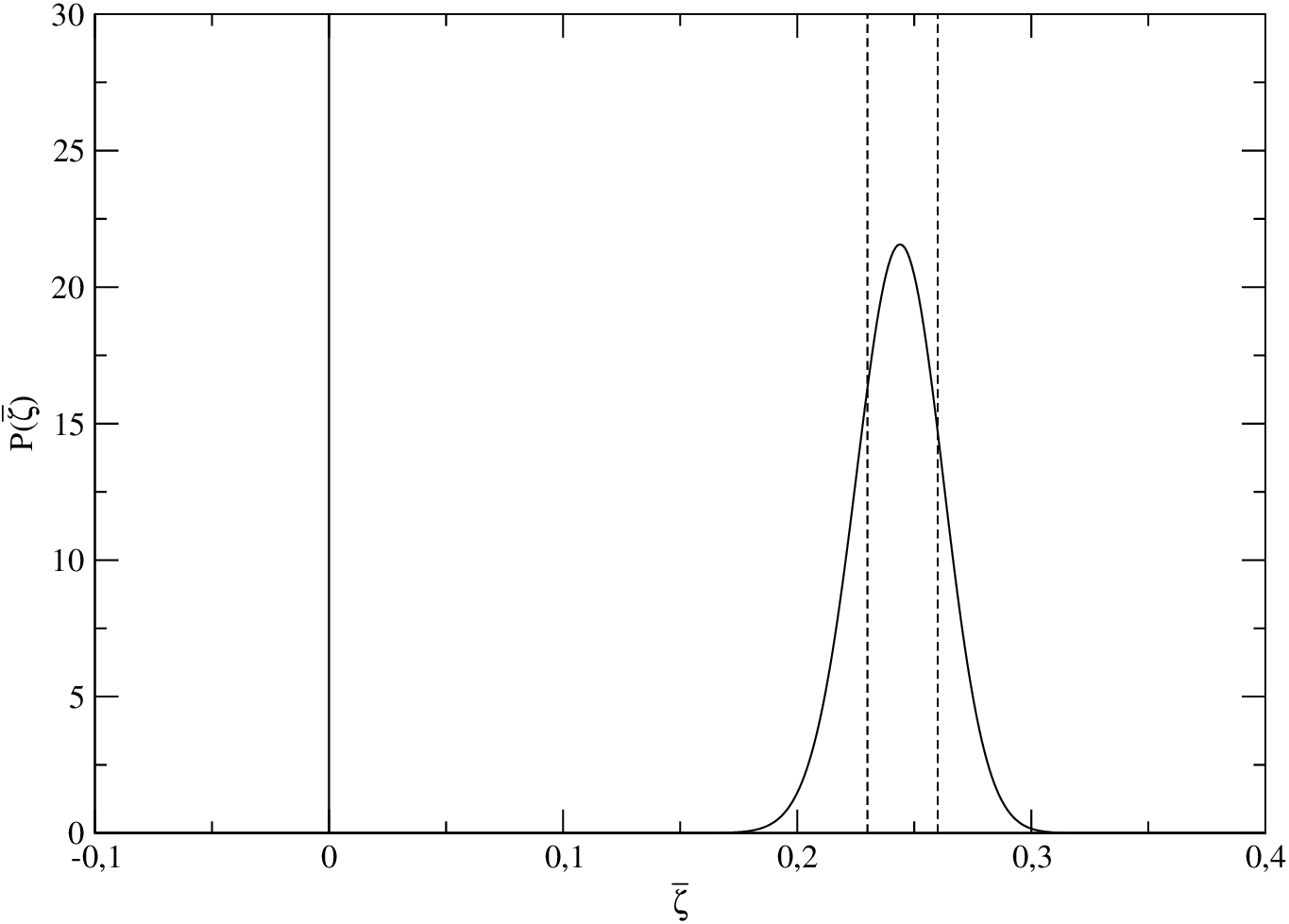}
\hspace{0.5cm}
\includegraphics[scale=0.21]{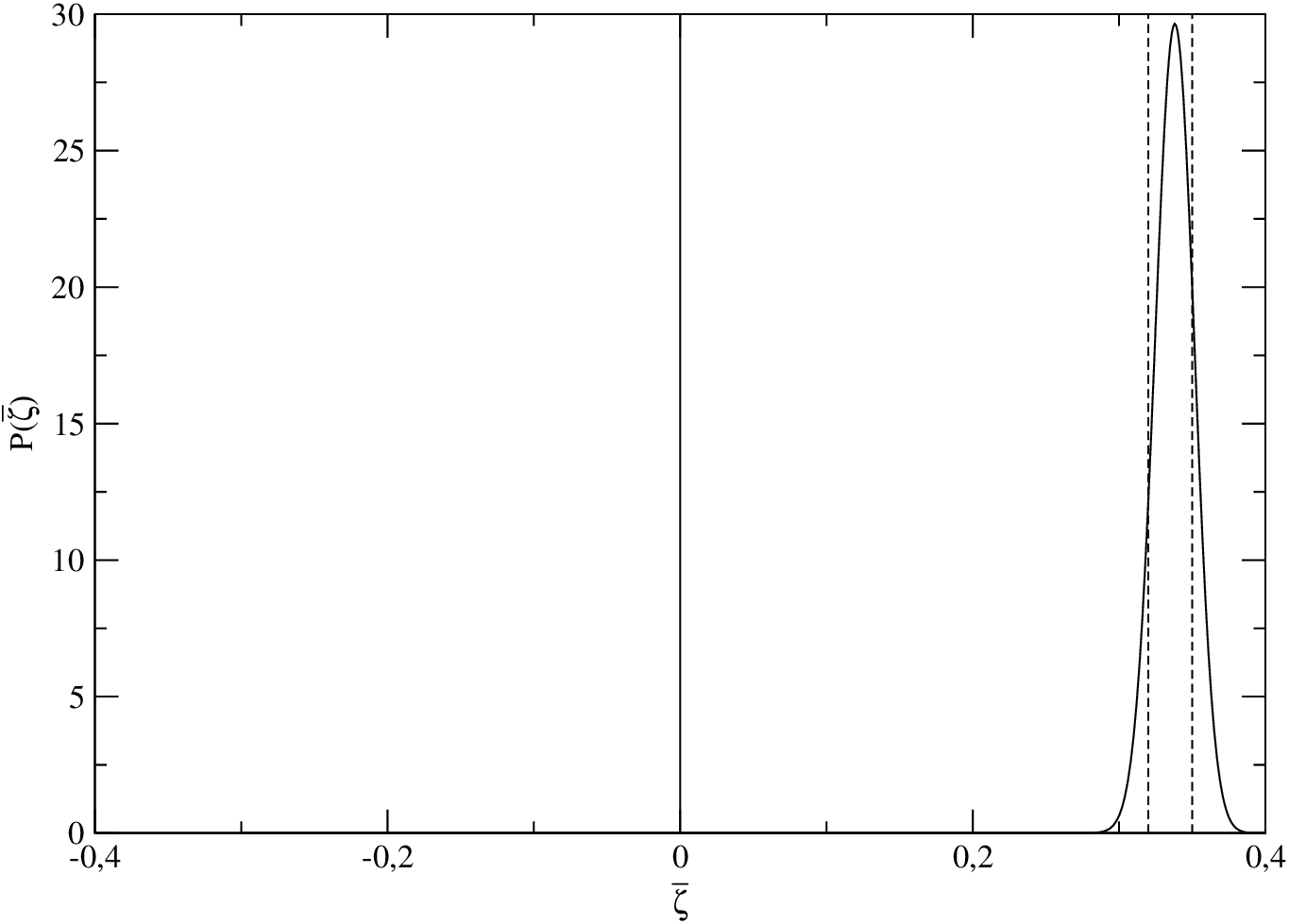}
\caption{P.d.f's concerning the cluster data divided in two sets, in
  increasing order of the redshift (below and above $z=0.05$). The
  data are compatible with one another and the results point to
  compatible values of $\bar\zeta $, at several standard deviations from a
  vanishing result. However, in the absence of weak lensing results
  the results point to a considerably higher value of the coupling
  compared with the previous ones.}
\label{figure2diaredshift}
\end{center}
\end{figure}

We can also check how robust are these results concerning the
redshift -- that is, we could ask whether evolution effects
are important. Hence, we divide the data in subsets in order of increasing
redshift -- and we do this only for the X-ray observations. The
results are displayed in (\ref{figure2diaredshift}).

\vspace{0.5cm}

The results for $\bar\zeta$ are, respectively, for smaller to larger
values of the redshift:
\bear
\bar\zeta_{z_{1}} &=& 0.24 \pm 0.01 \nonumber\\
\bar\zeta_{z_{2}}&=& 0.34 \pm 0.01 \; .
\label{redshiftresults}
\eear

Finally, some words are in order about choices made for
the available X-ray data for cluster. Not all clusters displayed in
\cite{girardif} are relaxed. Moreover, some data have different
results for different kinds of observations. In order to deal with
these differences we used two procedures. In one, we always considered
the data with largest radii. Since it is just a matter of random
choice, the actual result, if robust, should not depend on it. In the
second criterium, we considered the smallest
radii.
The results obtained using smaller and larger values of the radius are,
respectively:
\bear
\bar\zeta_{r_{1}}&=& 0.32 \pm 0.01 \nonumber\\
\bar\zeta_{r_{2}}&=& 0.30 \pm 0.01 \; .\label{radiusresults}
\eear
Hence, we conclude that for both criteria the results of our
analysis are very similar, which means that our constraints are
sufficiently robust with respect to these choices.

\section{Comparison with previous results}

In addition to the constraints obtained above, there are several other
results concerning the interaction of dark energy and dark matter.  A
possible transition of the Dark Energy equation of state has been
observed and an analysis of an explanation of such a transition in
terms of Dark Energy and Dark Matter interaction in the terms proposed
here has been performed in \cite{[12]}. Such an analysis leads to
results which can be translated to constraints in $\bar\zeta$ as 
\be
\bar\zeta = 0.18 \pm 0.18\quad .\label{trans} 
\ee 
That study has been
made using a holographic model for Dark Energy, but essential was the
interaction between the two dark sectors.

Furthermore, a study of the age of the old quasar APM 0879 + 5255 also
constrains the interaction \cite{[13]}. Indeed, such a quasar is too
old to be accomodated by the standard cosmology, and we have shown in
\cite{[13]} that an interaction leaking energy from the dark energy
into dark matter sector such that the coupling constant is given by
\be
\bar\zeta = 0.36 \pm 0.18\quad
\ee
naturally accomodates the age of the quasar being also compatible with
the age of the universe as given by CMB observations.

On the other hand, the small $\ell$ behavior of the CMB angular
spectrum \cite{huangshen} deviation from the standard model leads to
further constrains in the interaction as given by \cite{[13]} \be
\bar\zeta = 0.45 \pm 0.15\quad .  \ee

Furthermore, the interaction can also be modeled by an alternative
interacting field theory lagrangian, where Dark Energy is described by
a tachyonic lagrangian while the role of dark matter is played by a
fermionic field. A minimal interaction of the Yukawa type completes
the picture. Such a model is constrained by baryon acoustic
oscillations (BAO), lookback time, supernovae and CMB shift parameter
\cite{micheletti}. The effective coupling of both sectors, namely
fermionic (dark matter) and bosonic (dark energy) is constrained to
fulfill
\be
\bar\zeta = 0.17 \pm 0.09\quad .
\ee

Further observational data (once again taking into account supernovae,
shift parameter, BAO and galaxies ages estimates data) together with a
phenomenological model based on a two fluid interaction also points to
some (feeble) constraints \cite{feng}
\be
\bar\zeta = 0.01 \pm 0.04\quad ,
\ee
where we have considered only the most conservative result of that
analysis (in fact, the results present a weak dependence on the type of
interaction; we have taken the average, most conservative result).

Entirely theoretical arguments derived from a thermodynamical
analysis allows an estimation of the interaction parameter for the two fluid model as
\cite{wanglinpavonabd},
\be
\bar\zeta = 0.15 \pm 0.15\quad .
\ee

We also recall here a previous result concerning a smaller set of
clusters (contained in the total data set used in the present work):
\cite{ababrasw},
\be
\bar\zeta = 0.04 \pm 0.02\quad .
\ee

Our joint set of cluster observations led us to the constraint of Eq.
(\ref{joint}).  This is one way of viewing our result.  However,
considering the subsequent analysis (in particular the sample
variance), it may be fair to see the errors as much larger, possibly
of the order of magnitude of the highest difference in sub-averages.
Considering the highest spread to be $\delta\bar\zeta =
0.5\times(0.34-0.14)=0.10$, we take our results here as:
\be
\bar\zeta = 0.14 \pm 0.10\quad .\label{finres}
\ee

Even in this pessimistic case, we can infer that our result supports
the previous ones, i.e., that the interaction coupling parameter is a
small but positive and of the order of $0.1$, and almost three
standard deviations from vanishing.

\section{Conclusions}

We are fully aware that our lack of precise knowledge about the errors
arising from systematics could weaken our conclusions. We recall here
that our main work hypothesis is that the differences between the
various kinds of masses estimates come solely from the fact that they
do or do not take into account the interaction parameter. Obviously,
this assumption is not absolutely true, but represents an
approximation we use to replace the need to analyze any particular
systematic effects. Also, we have used in this work the best available
data sets well-suited for our purposes. Nevertheless, we still believe
that our analysis, combined with previous results, signals in a
coherent manner towards the possibility of the dark energy-dark matter
interaction.

The scenario in which dark energy and dark matter are in interaction
seems to be a strong physical possibility.  The interaction intensity
is a small constant, possibly larger than the fine structure constant,
in agreement with \cite{[17]}. The results are all consistent among
themselves, and we obtain the interesting conclusion that the
interaction constant is far from zero by several (possibly three)
standard deviations, which is a rather intriguing conclusion. The
positive value is also in agreement with the thermodynamical arguments
of \cite{wangpavon}, strengthening our conclusions. A few words should
be said about the possible influence of the interaction parameter on
the cosmic history (in particular, in the structure formation
process). The behavior of dark matter density as a function of time
can be rather different from the non interacting case depending on the
kind of interaction considered. We do not think that the proposed
interaction should be the same along the whole history since
decoupling, when a more robust suggestion has to be made. Since
usually the matter-radiation equality is supposed to occur, in the
standard model (e.g. in the non dark sector case) as well, the order
of magnitude of the cosmological time parameter should not get
strongly modified. (Notice that in the past, with interaction, dark
matter density was smaller compared to the noninteracting case.  Thus,
the matter density is bounded from below by the baryonic matter
density.)

\begin{acknowledgments}
This work has been supported by the Brazilian funding agencies FAPESP and
CNPq. We thank an anonymous referee for useful comments.
\end{acknowledgments}


\end{document}